\documentclass[twocolumn,showpacs,preprintnumbers,amsmath,amssymb,prb]{revtex4}
\usepackage{dcolumn}
\usepackage{bm}

\begin{document}

\title{The role of the two postulates of special relativity}

\author{Alfred Ziegler}
 \email{ziegler@th.physik.uni-frankfurt.de}
 \affiliation{
   Institut f{\"u}r theoretische Physik der Universit{\"a}t Frankfurt\\
   Max-von-Laue-Str.1, D-60438 Frankfurt, Germany
}

\date{\today}
\begin{abstract}
Students are often mystified by the reasoning that leads from the postulates of special relativity to the requirement of covariance. This is partly due to the lack of transparency resulting from the failure to clearly separate the roles the postulates of the constancy of the speed of light and of relativity play. Their roles are elucidated here by mainly geometric means. Finally some standard derivations found in textbooks are analyzed in order to sort out the basic physical ingredients.
\end{abstract}


\maketitle

\section{Introduction}\label{introduction}
It is well known that students find it difficult to grasp the essentials of the special theory of relativity. This is in spite of its mathematical simplicity unrivaled among the major subjects of physics. It has often been stated that it is absolutely necessary to get some feeling for relativistic phenomena first before one can proceed to the more formal parts as for instance the Lorentz transformation. Nevertheless many texts adopt a representation that proceeds to the
Lorentz transformation as quickly as possible in order to draw conclusions from it giving short shrift to the physical considerations inherent in the derivation. The emphasis of this article is not the – already well-documented – build-up of relativistic intuition but the treatment of the later part, viz. the derivation of the Lorentz transformation which is often presented in a confusing or misleading manner. In particular texts often fail to point out clearly which parts of the derivation rest on which of the two postulates on which the derivation is commonly based (relativity principle and constancy of the speed of light). I shall not dwell here on the topic of how to minimalize the assumptions required for the derivation, a topic on which an extensive literature with varied and often controversial answers already exists \cite{Test}
-- although interesting in itself this discussion is too academic for a first course in relativity --,
nor on how to make the derivation as simple as possible \cite{simple}. 
The aim of this article is rather to make as transparent as possible the way in which each postulate enters the derivation. Although
the emphasis is on the role of the two postulates it is also made clear where additional physical assumptions – unrelated to the postulates – required for the derivation of the Lorentz transformation, as for instance linearity, the definition of velocity etc. enter. Section \ref{results_from} addresses how the postulate of the constancy of the speed of light (or postulate of constancy in short) may be cast in mathematical form and which conclusions regarding the transformation of space and time may be drawn from this postulate alone, i.e. without invoking the principle of relativity. Section \ref{introducing_the} shows how the principle of relativity is used to demonstrate the invariance of length in the Minkowski metric, i.e. of $x^2-c^2t^2=l^2$ as opposed to the invariance of length of the null vectors $x^2-c^2t^2=0$ only as derived from the postulate of constancy. Making the transformation explicit requires the definition of velocity in addition. A treatment with all three spatial dimensions is sufficiently different from the one dimensional case to warrant a separate treatment in section \ref{the_case}. Section \ref{putting_the} shows how to proceed when one starts with the principle of relativity before implementing the postulate of constancy. Section \ref{direct_derivation} lists various derivations of the Lorentz transformations from textbooks and puts them into perspective (regarding sections \ref{results_from} and \ref{introducing_the}).

\section{Results from the postulate of constancy of the speed of light}\label{results_from}
For simplicity the discussion is restricted to one dimension (denoted $x$). The propagation of a light signal is described in two different frames of reference with their respective coordinates being given by $(x,t)$ and $(x',t')$. The postulate of constancy then takes the form $ds/dt=ds'/dt'=c$ where $ds$ ($ds'$) is the infinitesimal distance covered by the light in the infinitesimal time interval $dt$ ($dt´$). General as this formulation is, it is also pretty useless as starting point for the derivation of the transformation. Some preparatory measures have to be taken first. Unfortunately, they often go without saying.

\subsection{No change of direction}
First of all we restrict our discussion to light signals without reflection or, generally, change of direction. This converts $ds/dt=ds'/dt'=c$ into $\Delta s/\Delta t=\Delta s'/\Delta t'=c$. Because in one dimension $\Delta x/\Delta t=\pm c$, depending on whether the light signal is propagating in forward or backward direction, this translates into $\Delta x/\Delta t=\pm c\rightarrow\Delta x'/\Delta t'=\pm c$, both signs often being combined by writing $(\Delta x)^2-c^2(\Delta t)^2=0$. This form is still inconvenient.

\subsection{Linearity}
The coordinates are generally chosen so that seen from both frames of reference the light signal starts at the origin, i.e. at $x=x'=0$ and at the same time $t=t'=0$ (in other words: the clocks at the origin are synchronized at the start of the light signal). This simplifies the equation to
\begin{equation}\label{eq1} x^2-c^2t^2=x'^2-c^2t'^2=0. \end{equation}
This is not a matter of mere convenience but involves a physical assumption because equation (\ref{eq1}) does not imply
\begin{equation} (\Delta x)^2-c^2(\Delta t)^2=(\Delta x')^2-c^2(\Delta t')^2=0 \end{equation}
unless the transformation is linear. Otherwise
\begin{subequations}
\begin{equation} (\Delta x)'=(x_2-x_1)'\neq(x_2'-x_1') \end{equation}
\begin{equation} (\Delta t)'=(t_2-t_1)'\neq(t_2'-t_1') \end{equation}
\end{subequations}
This assumption reflects the homogeneity of space and time (the basis for the conservation of momentum and energy), i.e. that there are no distinguished points in either space or time (this argument is explained in more detail in  Resnick\cite{Resnick:1968}). We have finally reached the point where the postulate of constancy can be cast into mathematical form: we are looking for homogeneous linear transformations that leave the equation $x^2-c^2t^2=0$ invariant. This still leaves open various possibilities for the form of the transformation matrix (see appendix \ref{general_form}): the invariance of $x^2-c^2t^2=0$ means that the two diagonals of the $(x,ct)$-plane are mapped into each other collectively, not separately. Both diagonals could be mapped onto one, for instance, or the diagonals could be interchanged.

\subsection{No reversal of propagation}
Adopting as a further physical assumption that forward-propagating light signals remain forward-propagating in the other frame of reference (and backward-propagating signals backward-propagating) the diagonals turn into eigenvectors albeit with arbitrary eigenvalues $\alpha$ and $\beta$:
\begin{eqnarray} \lefteqn{x'^2-c^2t'^2=(x'-ct')(x'+ct')=}\nonumber\\
                 & & \alpha(x-ct)\beta(x+ct)=\alpha\beta(x^2-c^2t^2)=0 \end{eqnarray}
This is as far as one can get with the postulate of constancy alone.

\section{Introducing the principle of relativity}\label{introducing_the}
To put the principle of relativity into practice requires two independent steps.

\subsection{Step 1: equality of speed of relative motion}
The principle of relativity stipulates that the two frames of reference considered in the transformation be indistinguishable in their physical properties. This implies that observers in both frames of reference agree on the speed of their relative motion although not on the directions of motion which are opposite to each other, i.e. if the velocity is $v$ in one frame, it is $-v$ in the other. Transforming to another frame and back again must yield the identity transformation, which means in particular that the eigenvectors are carried into themselves. Since the eigenvalues depend on the only parameter of the frame transferred to, viz. the velocity $v$, one obtains for the combination of forward and backward transformation $\alpha(v)\alpha(-v)=\beta(v)\beta(-v)=1$.

\subsection{Step 2: equivalence of $v$ and $-v$}
On the other hand the operation of time reflection $t\rightarrow-t$ both interchanges the diagonals and carries $v$ into $-v$ in other words $\alpha(-v)=\beta(v)$ (see Greiner\cite{GreinerRafelski:1966} for more details).

\subsection{Synthesis}
Combining both equations leads to $1=\alpha(v)\alpha(-v)=\alpha(v)\beta(v)$. This is equivalent to the invariance of \emph{any} length in the Minkowski metric and thus the basis for covariance:
\begin{eqnarray} \lefteqn{x'^2-c^2t'^2=(x'-ct')(x'+ct')=}\nonumber\\
								 & &\alpha(x-ct)\beta(x+ct)=\alpha\beta(x^2-c^2t^2)=\nonumber\\
								 & & (x^2-c^2t^2).\end{eqnarray}
This general invariance does not in any way specify the function $\alpha(v)$. This again requires a physical argument.

\subsection{Definition of velocity}\label{definition_of}
It consists in the definition of the velocity $v$, which is quite natural because up to now $v$ is just a mathematical parameter devoid of any physical meaning. The origin of the system transfered to, $x'=0$, is moving by definition with velocity $v$ relative to the original system, i.e. $x=vt$ or $x-vt=0$. Due to the linearity of the transformation this leads to $x'\propto x-vt$ or $x'=\gamma(x-vt)$ with an as yet unspecified function $\gamma(v)$ respectively. The matrix that has as eigenvectors the diagonals with eigenvalues $\alpha$ and $\beta$ has the form
\begin{equation} \left( \begin{array}{cc} (\alpha+\beta)/2 & (\alpha-\beta)/2 \\ (\alpha-\beta)/2 & (\alpha+\beta)/2 \\ \end{array}\right) \end{equation}
(see appendix \ref{general_form_simplified}) leading to
\begin{equation} \left( \begin{array}{cc} (\alpha+\beta)/2 & (\alpha-\beta)/2 \\ (\alpha-\beta)/2 & (\alpha+\beta)/2 \\ \end{array}\right)\left(\begin{array}{c} x \\ ct\\ \end{array}\right)=\left(\begin{array}{c} x'\\ct'\\ \end{array}\right)\end{equation}
and
\begin{equation} \frac{\alpha+\beta}{2}x+\frac{\alpha-\beta}{2}ct=x' \end{equation}
from which by comparison with $x'=\gamma(x-vt)$
\begin{subequations}
\begin{equation}\begin{array}{cc} \gamma(v)=(\alpha+\beta)/2\textnormal{\quad} & {\displaystyle-\gamma(v)\frac{v}{c}=(\alpha-\beta)/2} \\ \end{array}\end{equation}
\begin{equation}\label{9b}\begin{array}{cc} \alpha=\gamma(v)(1-v/c)\textnormal{\quad} & \beta=\gamma(v)(1+v/c) \\ \end{array}\end{equation}
\end{subequations}
The transformation matrix then takes the form
\begin{equation}\gamma\left(\begin{array}{cc}1&-v/c\\-v/c&1\\\end{array}\right).\end{equation}
The product $\alpha\cdot\beta$ is related to $\gamma(v)$ by
\begin{equation}\alpha\cdot\beta=\gamma^2(v)(1-v^2/c^2).\end{equation}
Because $\alpha\cdot\beta=1$ one has $\gamma(v)=1/\sqrt{1-v^2/c^2}$ and one obtains the definite form of the transformation, the Lorentz transformation:
\begin{equation}\left(\begin{array}{c}x'\\ct'\\\end{array}\right)=\frac{1}{\sqrt{1-v^2/c^2}}\left(\begin{array}{cc}1&-v/c\\-v/c&1\\\end{array}\right)\left(\begin{array}{c}x\\ct\\\end{array}\right)\end{equation}

\subsection{Definition of velocity without principle of relativity}
One could as well reverse the introduction of velocity and the use of the principle of relativity. In this case one would still have relation (\ref{9b}) but no longer $\alpha\cdot\beta=1$. Arguing as above step 1 of the principle of relativity leads to
\begin{eqnarray}
\lefteqn{ \gamma(v)\left(\begin{array}{cc}1&-v/c\\-v/c&1\end{array}\right)\gamma(-v)\left(\begin{array}{cc}1&v/c\\v/c&1\end{array}\right)=}\nonumber\\
& &\gamma(v)\gamma(-v)\left(\begin{array}{cc}1-v^2/c^2&0\\0&1-v^2/c^2\end{array}\right)\nonumber\\
& & \stackrel{!}{=}\left(\begin{array}{cc}1&0\\0&1\end{array}\right)\end{eqnarray}
and $\gamma(v)\gamma(-v)\cdot(1-v^2/c^2)=1$, step 2 ($\gamma(v)=\gamma(-v)$) to $\gamma(v)=1/\sqrt{1-v^2/c^2}$ and
\begin{equation}\frac{1}{\sqrt{1-v^2/c^2}}\left(\begin{array}{cc}1&-v/c\\-v/c&1\end{array}\right)\end{equation}
for the transformation matrix.

\section{The case of three dimensions}\label{the_case}
Often all three dimensions are considered in the derivation of the Lorentz transformation. The argumentation is changed considerably thereby making the derivation rather complicated. To make use of the postulate of constancy one now has to determine all homogeneous linear transformations of fourdimensional space that leave the light cone $x^2+y^2+z^2-c^2t^2=0$ invariant. Whereas that was fairly easy for the $x - ct$-plane (either the transformation was degenerate projecting the whole plane onto one of the diagonals, or the diagonals were interchanged or eigenvectors) it is rather awkward for fourdimensional space. Also in contrast to the $x-ct$-plane where the diagonals were invariant separately no useful corresponding invariant subset of the light cone can be specified. Instead one adopts the more general requirement that the transformation reduces to the identical transformation for $v = 0$. The problem of classifying the invariance transformations is handled by invoking yet another physical argument: if the $x$-axis is the direction of motion $y'$ does not depend on $t$ (because of the assumption that the motion is in $x$-direction only) or on $x$ (because of the homogeneity of space). Thus $y'=\lambda y$ and $z'=\lambda z$. Because the invariance of $x^2+y^2+z^2-c^2t^2=0$ implies that
\begin{equation} x'^2+y'^2+z'^2-c^2t'^2=\mu(x^2+y^2+z^2-c^2t^2)\end{equation}
$\mu$ must be equal to $\lambda^2$ because for the special case
\begin{eqnarray} \lefteqn{x=x'=ct=ct'=0}\nonumber\\
                 & & y'^2+z'^2=\lambda^2(y^2+z^2)=\mu(y^2+z^2). \end{eqnarray}
$\mu$ plays the role of $\alpha\cdot\beta$ in the onedimensional case. The principle of relativity is then invoked to derive $\lambda=1$: because of the equivalence of $v$ and $-v$ $\lambda(v)=\lambda(-v)$; transforming back must lead to the identity $\lambda(v)^2=1$. The solution $\lambda=-1$ is not viable, because then even for $v=0$, i.e. identity of both frames one would obtain the paradoxical result $y=-y$. Thus $\lambda=1$. (A less formal and more intuitive form of this argument is the famous brush argument given in Taylor and Wheeler \cite{TaylorWheeler:1992}. There a moving train paints a line at a height of 1 m on a resting wall, i.e. $y'=1$ m, whereas on the wall is another brush that paints a line on the train at height $y=1$ m. Both lines must be at equal height, i.e. $y'=y=1$ m or $\lambda=1$, or otherwise one would come to the paradoxical conclusion that the lines are both below or above the other. Note that the position of the lines are objective because they do not change with time). The derivation then proceeds as in subsection \ref{definition_of} (with $y'=y$, $z'=z$ the homogeneous linear transformation of the $x$-$ct$-subspace is identical to the one derived in section \ref{results_from}).

\section{Putting the principle of constancy last}\label{putting_the}
In many texts the definition of velocity and the principle of relativity are implemented first before making use of the principle of constancy. This can be done in two ways: Setting $x'=a_{11}x+a_{12}ct=a_{11}(x-vt)$ (definition of velocity!) and $ct'=a_{21}x+a_{22}ct$ and using the light cone equations $x'^2-c^2t'^2=0$, $x^2-c^2t^2=0$ (for $x$, $t$ on the light cone) or the equivalent equation $x'^2-c^2t'^2=\mu(x^2-c^2t^2)$ (for arbitrary $x$, $t$) one obtains
\begin{eqnarray} \lefteqn{x^2(a_{11}^2-a_{21}^2-\mu)-2xct(a_{11}^2v/c+a_{21}a_{22})}\nonumber\\
                 & & +c^2t^2(\mu-a_{22}^2+a_{11}^2v^2/c^2)=0. \end{eqnarray}
The coefficients of $x^2$, $xct$, and $c^2t^2$ may now be set equal to zero separately because $x$ and $ct$ are independent. Solving these equations yields the Lorentz transformation, after the relativity principle has been used to set $\mu=1$ (which could as well have been done at the outset). This argument is not applicable in the onedimensional case because there $x=\pm ct$ is not independent of $ct$. The resulting three equations for the three unknowns $a_{11}$, $a_{12}$, $a_{22}$ may be solved to obtain either the Lorentz transformation or the Lorentz transformation combined with a point reflection which is then ruled out because it does not lead to the identity transformation in the limit $v=0$. This procedure is akin to the the derivation of the general homogeneous linear transformation given in appendix \ref{general_form} with the difference that the degenerate transformations have already been eliminated by the condition $y'=\lambda y$, $z'=\lambda z$. Alternatively one could have started with the equations
\begin{equation}\label{ansatz}x'=\gamma(x-vt)\quad x=\gamma(x'+vt').\end{equation}
They already contain the principle of relativity (the form of the second equation embodies step 1, the use of the same $\gamma$ step 2). Using the light cone equations, i.e. $x=ct$, $x'=ct'$ results in
\begin{equation} ct'=\gamma(ct-vt)\quad ct=\gamma(ct'+vt')\end{equation}
and $\gamma=1/\sqrt{1-v^2/c^2}$.

\section{Direct derivation of time dilation and length contraction}\label{direct_derivation}
Both effects may be derived without reference to the Lorentz transformation. Still it is important to mark where the principle of relativity enters. One could for instance use a light signal that transverses a moving train from front to back where it is reflected back to the front. The propagation of this signal is described both in the frame of the train and the frame of the platform. This setup is the most natural one to derive (both quantitatively and qualitatively, see Hobson \cite{Hobson:2003}) the three principal kinematical results of the special theory of relativity: time dilation, lenght contraction, and relativity of simultaneity. The problem is that one deals with the effects of time dilation and length contraction simultaneously. This is the reason why one ususally starts with a light signal transverse to the direction of motion to derive the time dilation (see appendix \ref{standard_derivation}): there is no length contraction in the transverse direction. This is usually taken to be granted or as plain obvious. Actually as pointed out in the previous section it is an embodiment of the relativity principle. Using the relativity principle directly it is possible to deal with time dilation and length contraction at the same time. The postulate of constancy leads to $t'=2l/c$ for the time measured in the train ($l=$ length of the train at rest) and
\begin{equation}t=\frac{l'}{c-v}+\frac{l'}{c+v}=\frac{2l'c}{c^2-v^2}\end{equation}
for the time measured from the platform. Again we need the principle of relativity to reduce the unknowns from four to two: $l/t=l'/t'=v$ (because both observers agree on the velocity relative to each other). The result is
\begin{equation}\frac{t'}{t}=\frac{2l}{c}\cdot\frac{c^2-v^2}{2l'c}=\frac{l}{l'}(1-\frac{v^2}{c^2})=\frac{t}{t'}(1-\frac{v^2}{c^2})\end{equation}
from which $t'/t=\sqrt{1-v^2/c^2}$.

\section{A survey of textbooks}\label{a_survey}
In the following an overview is given over representative derivations found in textbooks. This is done for the purpose of pointing out where the postulates make their appearance in the derivation because – as was mentioned above – their respective roles are often not very transparent. The overview is broken down in correspondence with the items listed in section \ref{results_from} and \ref{introducing_the}.

\subsection{Linearity}
The majority of textbooks sees no need to justify the linearity of the transformation \cite{Westphal:1968,Giancoli:2000,TaylorWheeler:1992,Tipler:2003,KittelKnightRuderman:1968,Goldstein:2003,BallifDibble:1969,AlonsoFinn:1966,Sartori:1996,Joos:1987}. When given it is derived either from homogeneity of time and space \cite{Jackson:1998,Rindler:2001,LandauLifshitz:1951,Resnick:1968,Ford:1977,BeckerSauter:1982} or from Newton\textquoteright s first law which requires uniform and thus straight motion to remain uniform and straight in another system of reference \cite{GreinerRafelski:1966,French:1968,Moller:1972}. Resnick \cite{Resnick:1968} is the most explicit text in this respect. Actually a lot more ``obvious'' statements regarding the possible choices of coordinates have to be demonstrated which among the texts listed here are only given in Rindler \cite{Rindler:2001}.

\subsection{Invariance of the transverse coordinates}
This has of course to be demonstrated only when using three dimensions which virtually all texts do. Surprisingly many texts just present it as obvious. The ones that take the effort \cite{Rindler:2001,LandauLifshitz:1951,TaylorWheeler:1992,BallifDibble:1969,Resnick:1968,Ford:1977,French:1968,Moller:1972,BeckerSauter:1982} more or less use variants of the brush argument. Among the more detailed explanations are Taylor and Wheeler \cite{TaylorWheeler:1992}, Ballif-Dibble \cite{BallifDibble:1969} and Resnick \cite{Resnick:1968}.

\subsection{Principle of relativity}
None of the texts reviewed here use the geometric approach of sections \ref{results_from} and \ref{introducing_the} which makes it often difficult to perceive clearly where step 1 and step 2 enter the derivation. Since nearly all the texts opt for the sequence ``principle of relativity $\rightarrow$ principle of constancy'' they fall mainly into two categories. Either they make full use of the result $\mu=1$ - basically the invariance of the transverse coordinates – as in the first part of section \ref{putting_the}\cite{Jackson:1998,TaylorWheeler:1992,KittelKnightRuderman:1968,Goldstein:2003,AlonsoFinn:1966,Resnick:1968,Joos:1987,Moller:1972}.  This result is derived by the relativity principle. Only few texts go beyond the brush argument to point out where step 1 and 2 enter. A succinct but good example is Jackson \cite{Jackson:1998}: $\mu(v)\cdot\mu(-v)=1$ or otherwise the combination of forward and backward transformation differs from the identity (step 1) and $\mu(v)$ must be independent of the sign of $v$ because it represents a scale change in the direction transverse to the direction of motion (step 2). Or alternatively the texts proceed as in the second part of section \ref{putting_the}\cite{Westphal:1968,Giancoli:2000,Rindler:2001,GreinerRafelski:1966,Tipler:2003,BallifDibble:1969,Sartori:1996,French:1968,BeckerSauter:1982}. Only Greiner \cite{GreinerRafelski:1966}, Rindler \cite{Rindler:2001} and Becker-Sauter \cite{BeckerSauter:1982} (if briefly) explain the reasons for the ansatz (\ref{ansatz}) where Greiner is the most explicit of these texts.

\subsection{Sign ambiguity}
The solutions of the equations obtained by setting the coefficients equal to zero contain sign ambiguities. The sign has to be chosen so that one obtains the identity for $v=0$. This ambiguity is addressed by very few texts \cite{Jackson:1998}.

\section{Summary}
The individual steps in the derivation of the Lorentz transformation have been pointed out with a geometric derivation as reference. The separate roles of the postulates of constancy and relativity have been clearly delineated. It was emphasized that the application of the principle of relativity comprises two steps: the combination of forward and backward transformation must yield the identity transformation (step 1) and forward and backward transformation are identical apart from the different sign of the velocity (step 2). Standard textbook derivations have been put into perspective relative to the reference derivation.

\appendix
\section{General form of the transformation matrix (simplified derivation)}\label{general_form_simplified}
The general form for a homogeneous linear transformation in two dimensions is:
\begin{subequations}
\begin{equation} x_0'=a_{11}x_0+a_{12}x_1 \end{equation}
\begin{equation} x_1'=a_{21}x_0+a_{22}x_1 \end{equation}
\end{subequations}
from which
\begin{equation} x_0'+\sigma x_1'=(a_{11}+\sigma a_{21})x_0+(a_{12}+\sigma a_{22})x_1 \end{equation}
Assuming that both diagonals given by $x_0+\sigma x_1=0$ or $x_0=-\sigma x_1$ are mapped to their counterparts, i.e. principal to principal diagonal and secondary to secondary diagonal, one obtains
\begin{equation} x_0'+\sigma x_1'=-\sigma(a_{11}+\sigma a_{21})x_1+(a_{12}+\sigma a_{22})x_1\stackrel{!}{=}0 \end{equation}
Because this equation must hold for both signs it follows that $a_{11}=a_{22}=:d$ and $a_{12}=a_{21}=:o$. The eigenvalues are $\alpha=d+o$ and $\beta=d-o$. Correspondingly the matrix elements are $d=(\alpha+\beta)/2$ and $o=(\alpha-\beta)/2$.

\section{General form of the transformation matrix}\label{general_form}
The general form for a homogeneous linear transformation in two dimensions is:
\[ \begin{array}{cc}
x'_0=a_{11}x_0+a_{12}x_1 & x'^2_0=a_{11}^2x_0^2+a_{12}^2x_1^2+2a_{11}a_{12}x_0x_1 \\
x'_1=a_{21}x_0+a_{22}x_1 & x'^2_1=a_{21}^2x_0^2+a_{22}^2x_1^2+2a_{21}a_{22}x_0x_1 \\
\end{array}\]
Assuming that the Minkowski length is zero in both coordinate systems leads to
\begin{eqnarray*} \lefteqn{x'^2_0-x'^2_1=(a_{11}^2-a_{21}^2)x_0^2-(a_{22}^2-a_{12}^2)x_1^2+}\\
& & 2(a_{11}a_{12}-a_{21}a_{22})x_0x_1\stackrel{!}{=}0\end{eqnarray*}
Writing $x_0^2-x_1^2=0$ in the form $x_1=\sigma x_0$ ($\sigma=\pm1$) one obtains
\begin{eqnarray*} \lefteqn{x'^2_0-x'^2_1=(a_{11}^2-a_{22}^2-a_{21}^2+a_{12}^2)x_0^2+}\\
& & 2(a_{11}a_{12}-a_{21}a_{22})x_0^2\sigma\stackrel{!}{=}0\end{eqnarray*}
Because these relations must hold for arbitrary $x_0$ and $\sigma$ $a_{11}^2-a_{22}^2=a_{21}^2-a_{12}^2$ and $a_{11}a_{12}=a_{21}a_{22}$ or by renaming the matrix elements
\[\left(\begin{array}{cc}a_{11}&a_{12}\\a_{21}&a_{22}\\\end{array}\right)=\left(\begin{array}{cc}d_1&o_1\\o_2&d_2\\\end{array}\right)\]
$d_1^2-d_2^2=o_2^2-o_1^2$ and $d_1o_1=d_2o_2$ from which
\[d_1^2-d_2^2=\frac{d_1^2}{d_2^2}o_1^2-o_1^2\Rightarrow d_2^4-(d_1^2+o_1^2)d_2^2+d_1^2o_1^2=0\]
This quadratic equation has solutions $d_2^2=d_1^2$ and $d_2^2=o_1^2$ as can be easily seen by Vieta's theorem.

a) $d_2^2=d_1^2\Rightarrow o_2^2=o_1^2$ and $d_2=\sigma d_1\Rightarrow o_2=\sigma o_1$. The resulting equation is
\[\left(\begin{array}{cc}d&o\\\sigma o&\sigma d\\\end{array}\right)\left(\begin{array}{c}1\\\pm1\\\end{array}\right)=\left(\begin{array}{c}d\pm o\\\sigma(o\pm d)\\\end{array}\right)=(d\pm o)\left(\begin{array}{c}1\\\pm\sigma\\\end{array}\right)\]
The choice of $\sigma = 1$ leaves the diagonals invariant separately whereas $\sigma=-1$ interchanges them.

b) $d_2^2=o_1^2\Rightarrow d_1^2=o_2^2$ and $d_2=\sigma o_1\Rightarrow d_1=\sigma o_2$. The resulting eigenvalue equation is
\[\left(\begin{array}{cc}\sigma o_2&o_1\\o_2&\sigma o_1\\\end{array}\right)\left(\begin{array}{c}1\\\pm1\\\end{array}\right)=\left(\begin{array}{c}\sigma o_2\pm o_1\\o_2\pm\sigma o_1\\\end{array}\right)=(\sigma o_2\pm o_1)\left(\begin{array}{c}1\\\sigma\\\end{array}\right)\]
The choice of $\sigma=1$ maps both diagonals to the principal diagonal whereas $\sigma=-1$ maps them to the secondary diagonal.

Only case a) with $\sigma=1$ preserves the forward and backward propagation. As above the eigenvalues in this case are $\alpha=d+o$ and $\beta=d-o$ and the matrix elements are $d=(\alpha+\beta)/2$ and $o=(\alpha-\beta)/2$.

\section{Standard derivation of time dilation}\label{standard_derivation}
A light signal in a moving train propagates transverse to the direction of motion from the floor to the ceiling, where it is reflected, and back. The distance traveled is $2h$ with $h$ being the height of the train. Seen from the platform the distance traveled by the light is longer because while moving towards the ceiling the light also travels forward by the amount $vt$, where $t$ is the time measured in the frame of the platform. The total distance traveled is easily obtained by the Pythagorean theorem as $2\sqrt{h^2+v^2t^2}=2ct$ where the equality sign reflects the postulate of constancy ($t$ is taken here as the time for the trip from the floor to the ceiling). Solving for $t$ one obtains $t=\sqrt{h^2/(c^2-v^2)}$. The speed of light must be the same in the frame of the train. Because the distance covered is shorter the inevitable conclusion is that less time has passed. Denoting the time passed in the train during the passage of the light from ceiling to bottom by $t'$, i.e. $h=ct'$, one has $t=\sqrt{c^2t'^2/(c^2-v^2)}$ or $t=t'/\sqrt{1-v^2/c^2}$.

\newpage
\bibliographystyle{apsrev}
\bibliography{dokument}

\begin{thebibliography}{22}
\expandafter\ifx\csname natexlab\endcsname\relax\def\natexlab#1{#1}\fi
\expandafter\ifx\csname bibnamefont\endcsname\relax
  \def\bibnamefont#1{#1}\fi
\expandafter\ifx\csname bibfnamefont\endcsname\relax
  \def\bibfnamefont#1{#1}\fi
\expandafter\ifx\csname citenamefont\endcsname\relax
  \def\citenamefont#1{#1}\fi
\expandafter\ifx\csname url\endcsname\relax
  \def\url#1{\texttt{#1}}\fi
\expandafter\ifx\csname urlprefix\endcsname\relax\def\urlprefix{URL }\fi
\providecommand{\bibinfo}[2]{#2}
\providecommand{\eprint}[2][]{\url{#2}}

\bibitem[{Tes()}]{Test}
\bibinfo{note}{Only a partial list can be given here:\\Jean-Marie
  L{\'e}vy-LeBlond, "One more derivation of the Lorentz transformation",
  Am.J.Phys. \textbf{44} 271-277 (1976)\\Bernhard Rothenstein, George Eckstein,
  "Lorentz transformations directly from the speed of light", Am.J.Phys.
  \textbf{63} 1150 (1995)\\H.M. Schwartz, "Deduction of the general Lorentz
  transformations from a set of necessary assumptions", Am.J.Phys. \textbf{52}
  346-350 (1984)\\A.R. Lee, T.M. Kalotas, "Lorentz transformations from the
  first postulate", Am.J.Phys. \textbf{43} 434-437 (1977)\\N.D. Mermin,
  "Relativistic addition of velocities directly from the constancy of the
  velocity of light", Am.J.Phys. \textbf{51} 1130-1131 (1983)}.

\bibitem[{sim()}]{simple}
\bibinfo{note}{See e.g.:\\Jean-Marie L{\'e}vy, "A simple derivation of the
  Lorentz transformation and of the accompanying velocity and accelaration
  changes", Am.J.Phys. \textbf{75} 615-618 (2007)\\W.N.Mathews,Jr.,
  "Relativistic velocity and acceleration transformations from thought
  experiments", Am.J.Phys. \textbf{73} 45-51, and references therein\\ Alan
  Macdonald, "Derivation of the Lorentz transformation", Am.J.Phys. \textbf{49}
  493 (1981)\\ David Park, "Derivation of the Lorentz transformation from
  gedanken experiments", Am.J.Phys. \textbf{42} 909-910 (1974)}.

\bibitem[{\citenamefont{Resnick}(1968)}]{Resnick:1968}
\bibinfo{author}{\bibfnamefont{R.}~\bibnamefont{Resnick}},
  \emph{\bibinfo{title}{Introduction to special relativity}}
  (\bibinfo{publisher}{John Wiley}, \bibinfo{address}{New York},
  \bibinfo{year}{1968}).

\bibitem[{\citenamefont{Greiner and Rafelski}(1966)}]{GreinerRafelski:1966}
\bibinfo{author}{\bibfnamefont{W.}~\bibnamefont{Greiner}} \bibnamefont{and}
  \bibinfo{author}{\bibfnamefont{J.}~\bibnamefont{Rafelski}},
  \emph{\bibinfo{title}{Spezielle Relativit{\"a}tstheorie}}
  (\bibinfo{publisher}{Harri Deutsch}, \bibinfo{address}{Frankfurt},
  \bibinfo{year}{1966}).

\bibitem[{\citenamefont{Taylor and Wheeler}(1992)}]{TaylorWheeler:1992}
\bibinfo{author}{\bibfnamefont{E.}~\bibnamefont{Taylor}} \bibnamefont{and}
  \bibinfo{author}{\bibfnamefont{J.}~\bibnamefont{Wheeler}},
  \emph{\bibinfo{title}{Spacetime Physics}} (\bibinfo{publisher}{W.H. Freeman},
  \bibinfo{address}{San Francisco}, \bibinfo{year}{1992}).

\bibitem[{\citenamefont{Hobson}(2003)}]{Hobson:2003}
\bibinfo{author}{\bibfnamefont{A.}~\bibnamefont{Hobson}},
  \emph{\bibinfo{title}{Physics, Concepts and Connections}}
  (\bibinfo{publisher}{Prentice Hall}, \bibinfo{address}{Upper Saddle River},
  \bibinfo{year}{2003}).

\bibitem[{\citenamefont{Westphal}(1968)}]{Westphal:1968}
\bibinfo{author}{\bibfnamefont{W.}~\bibnamefont{Westphal}},
  \emph{\bibinfo{title}{Physik, eine Einf{\"u}hrung}}
  (\bibinfo{publisher}{Springer}, \bibinfo{address}{New York},
  \bibinfo{year}{1968}).

\bibitem[{\citenamefont{Giancoli}(2000)}]{Giancoli:2000}
\bibinfo{author}{\bibfnamefont{D.}~\bibnamefont{Giancoli}},
  \emph{\bibinfo{title}{Physics for Scientists and Engineers}}
  (\bibinfo{publisher}{Prentice Hall}, \bibinfo{address}{Upper Saddle River},
  \bibinfo{year}{2000}).

\bibitem[{\citenamefont{Tipler}(2003)}]{Tipler:2003}
\bibinfo{author}{\bibfnamefont{P.}~\bibnamefont{Tipler}},
  \emph{\bibinfo{title}{Modern Physics}} (\bibinfo{publisher}{W.H. Freeman},
  \bibinfo{address}{San Francisco}, \bibinfo{year}{2003}).

\bibitem[{\citenamefont{Kittel et~al.}(1968)\citenamefont{Kittel, Knight, and
  Ruderman}}]{KittelKnightRuderman:1968}
\bibinfo{author}{\bibfnamefont{C.}~\bibnamefont{Kittel}},
  \bibinfo{author}{\bibfnamefont{W.}~\bibnamefont{Knight}}, \bibnamefont{and}
  \bibinfo{author}{\bibfnamefont{M.}~\bibnamefont{Ruderman}},
  \emph{\bibinfo{title}{Berkeley Course in Physics, Vol.1}}
  (\bibinfo{publisher}{McGraw-Hill}, \bibinfo{address}{New York},
  \bibinfo{year}{1968}).

\bibitem[{\citenamefont{Goldstein}(2003)}]{Goldstein:2003}
\bibinfo{author}{\bibfnamefont{H.}~\bibnamefont{Goldstein}},
  \emph{\bibinfo{title}{Classical Mechanics}}
  (\bibinfo{publisher}{Addison-Wesley}, \bibinfo{address}{Reading},
  \bibinfo{year}{2003}).

\bibitem[{\citenamefont{Ballif and Dibble}(1969)}]{BallifDibble:1969}
\bibinfo{author}{\bibfnamefont{J.}~\bibnamefont{Ballif}} \bibnamefont{and}
  \bibinfo{author}{\bibfnamefont{W.}~\bibnamefont{Dibble}},
  \emph{\bibinfo{title}{Conceptual Physics}} (\bibinfo{publisher}{John Wiley},
  \bibinfo{address}{New York}, \bibinfo{year}{1969}).

\bibitem[{\citenamefont{Alonso and Finn}(1980)}]{AlonsoFinn:1966}
\bibinfo{author}{\bibfnamefont{M.}~\bibnamefont{Alonso}} \bibnamefont{and}
  \bibinfo{author}{\bibfnamefont{E.}~\bibnamefont{Finn}},
  \emph{\bibinfo{title}{Fundamental University Physics, Vol.1}}
  (\bibinfo{publisher}{Addison-Wesley}, \bibinfo{address}{Reading},
  \bibinfo{year}{1980}).

\bibitem[{\citenamefont{Sartori}(1996)}]{Sartori:1996}
\bibinfo{author}{\bibfnamefont{L.}~\bibnamefont{Sartori}},
  \emph{\bibinfo{title}{Understanding relativity}} (\bibinfo{publisher}{Univ.
  of California Press}, \bibinfo{address}{Berkeley}, \bibinfo{year}{1996}).

\bibitem[{\citenamefont{Joos}(1987)}]{Joos:1987}
\bibinfo{author}{\bibfnamefont{G.}~\bibnamefont{Joos}},
  \emph{\bibinfo{title}{Theoretical Physics}} (\bibinfo{publisher}{Dover
  Publications}, \bibinfo{address}{Mineola}, \bibinfo{year}{1987}).

\bibitem[{\citenamefont{Jackson}(1998)}]{Jackson:1998}
\bibinfo{author}{\bibfnamefont{J.}~\bibnamefont{Jackson}},
  \emph{\bibinfo{title}{Electrodynamics}} (\bibinfo{publisher}{Wiley},
  \bibinfo{address}{New York}, \bibinfo{year}{1998}).

\bibitem[{\citenamefont{W.Rindler}(2001)}]{Rindler:2001}
\bibinfo{author}{\bibnamefont{W.Rindler}}, \emph{\bibinfo{title}{Relativity}}
  (\bibinfo{publisher}{Oxford University Press}, \bibinfo{address}{Oxford},
  \bibinfo{year}{2001}).

\bibitem[{\citenamefont{Landau and Lifshitz}(1995)}]{LandauLifshitz:1951}
\bibinfo{author}{\bibfnamefont{L.}~\bibnamefont{Landau}} \bibnamefont{and}
  \bibinfo{author}{\bibfnamefont{E.}~\bibnamefont{Lifshitz}},
  \emph{\bibinfo{title}{The Classical Theory of Fields}}
  (\bibinfo{publisher}{Elsevier}, \bibinfo{address}{Amsterdam},
  \bibinfo{year}{1995}).

\bibitem[{\citenamefont{Ford}(1977)}]{Ford:1977}
\bibinfo{author}{\bibfnamefont{K.}~\bibnamefont{Ford}},
  \emph{\bibinfo{title}{Classical and Modern Physics, Vol.3}}
  (\bibinfo{publisher}{John Wiley}, \bibinfo{address}{New York},
  \bibinfo{year}{1977}).

\bibitem[{\citenamefont{Becker and Sauter}(1982)}]{BeckerSauter:1982}
\bibinfo{author}{\bibfnamefont{R.}~\bibnamefont{Becker}} \bibnamefont{and}
  \bibinfo{author}{\bibfnamefont{F.}~\bibnamefont{Sauter}},
  \emph{\bibinfo{title}{Electromagnetic fields and interactions}}
  (\bibinfo{publisher}{Dover publications}, \bibinfo{address}{Mineola},
  \bibinfo{year}{1982}).

\bibitem[{\citenamefont{French}(1989)}]{French:1968}
\bibinfo{author}{\bibfnamefont{A.}~\bibnamefont{French}},
  \emph{\bibinfo{title}{Relativity}} (\bibinfo{publisher}{MIT Press},
  \bibinfo{address}{Boston}, \bibinfo{year}{1989}).

\bibitem[{\citenamefont{M{\o}ller}(1972)}]{Moller:1972}
\bibinfo{author}{\bibfnamefont{C.}~\bibnamefont{M{\o}ller}},
  \emph{\bibinfo{title}{Theory of Relativity}} (\bibinfo{publisher}{Oxford
  University Press}, \bibinfo{address}{Oxford}, \bibinfo{year}{1972}).

\end{thebibliography}

\end{document}